\documentclass[twocolumn]{emulateapj}

\usepackage{epsfig}
\usepackage{amsmath}
\usepackage{graphicx}
\usepackage{rotating}
\usepackage{tabularx}
\usepackage{amssymb}
\usepackage{subfigure}
\usepackage{epstopdf}

\shorttitle{}

\begin{document}

\title{Imaging Redshift Estimates for {\it Fermi} BL Lacs}
\author{Matt Stadnik \& Roger W. Romani\altaffilmark{1}}
\affil{Dept. of Physics/KIPAC, Stanford University, Stanford, CA 94305-4060}
\altaffiltext{1}{Visiting Astronomer, Kitt Peak National and Cerro Tololo Inter-American 
Observatories, which are operated by the Association of Universities for Research in Astronomy 
(AURA) under cooperative agreement with the National Science Foundation. 
The WIYN observatory is a joint facility of the University of Wisconsin-Madison, 
Indiana University, Yale University, and the National Optical Astronomy Observatory.
The SOAR telescope is a joint project of: Conselho Nacional de Pesquisas Cientificas e 
Tecnologicas CNPq-Brazil, The University of North Carolina at Chapel Hill, Michigan State 
University, and the National Optical Astronomy Observatory.
}

\email{mstadnik@stanford.edu, rwr@astro.stanford.edu \qquad\qquad\qquad}

\begin{abstract}
We have obtained WIYN and SOAR $i^{\prime}$ images of BL Lacertae objects (BL Lacs)
and used these to detect or constrain the flux of the host galaxy. Under common 
standard candle assumptions these data provide estimates of, or lower bounds on, the redshift. 
Our targets are a set of flat-spectrum radio counterparts of high flux {\it Fermi} Large Area 
Telescope (LAT) sources, with sensitive spectral observations showing
them to be continuum-dominated BL Lacs. In this sample 5 of 11 BL Lacs yielded significant host 
detections, with standard candle redshifts $z=0.13-0.58$. Our estimates and lower bounds
are generally in agreement with other redshifts estimates, although our $z=0.374$ estimate
for J0543$-$5532 implies a significantly sub-luminous host.
\end{abstract}


\section{Introduction}

BL Lac objects are an important, extreme sub-class of AGN. The large numbers of BL Lacs
being detected in the gamma-rays by the {\it Fermi} LAT \citep{2lac} show that these blazars
are a dominant contributor to the GeV sky. This flux-limited gamma-ray sample also provides 
a unique opportunity to probe the evolution of the BL Lac phenomenon over cosmic time
\citep{aet14}.  BL Lac evolution is controversial, with various authors finding
negative \citep[e.g.][]{rec00}, positive \citep[e.g.][]{mc13} or negligible evolution \citep[e.g.][]{cac02}.
Much of this uncertainty stems from the difficulty in determining redshifts;
for many BL Lacs the high jet-associated continuum flux 
dominates emission from any broad-line region and results in very low equivalent width 
for galactic host absorption features. Despite intensive spectroscopic campaigns
\citep{shaw13b} this often precludes direct BL Lac redshift measurement,
challenging their utility for population and evolution studies \citep{aet14}. Thus alternate methods
of estimating or constraining the source redshift can be quite valuable \citep{shaw13a}.

In this paper, we image {\it Fermi} BL Lacs to search for host flux in the wings of the nuclear
point spread function (PSF). Our targets are drawn from the `2LAC' \citep{2lac} LAT blazars for which sensitive
spectroscopy, reported in \citet{shaw13b}, shows them to be continuum-dominated BL Lacs
at very high significance, but was not able to determine a redshift. Since hosts are increasingly
difficult to detect at high redshift, we excluded BL Lacs whose spectra showed 
intervening absorbers requiring $z>0.7$ or excluded continuum flux from hosts at lower $z$; this
left 226 targets. We observed objects from this list when conditions and gaps in 
other observing programs allowed.

It has been argued that BL Lac hosts are nearly uniform giant ellipticals with
$M_R$ = $-$22.5 \citep{sba05}. Thus several authors have used such imaging to constrain
the source redshift \citep[e.g.][]{sba05,meisner1,nilet12}. One can also search
for the elliptical flux in high S/N spectra; when not detected this provides
a lower limit on the redshift \citep[e.g.][]{sba06,shaw13b,sand13}.
Conversely, when direct spectroscopic redshifts
measurements are in hand (or later become available), the flux measurements can be used
to test the standard candle hypothesis and the evolution of the BL Lac hosts. Here, we
follow the method described by \citet{meisner1}, although we use an updated
host magnitude calibration derived from absorption lines features measured in a deep 
spectroscopic survey \citep{shaw13a}.

\section{Observation and Data Reduction}

	Our typical BL Lac has a total magnitude of $i^\prime \sim 15-18$. We wish to
model our individual image point spread function (PSF), following the wings to well 
below the sky brightness, and to test the PSF model against a number of stars
with flux comparable to that of the BL Lac core. Thus we need moderate field imaging
with good natural seeing. For this program we used the Mini-Mosaic (MiniMo) camera
at the WIYN 3.6m telescope at Kitt Peak National Observatory and the 
SOAR Optical Imager (SOI) at the SOAR 4.2m telescope on Cerro Pachon. Individual
exposures ($t_{exp} = 180$\,s$-600$\,s) were adjusted to minimize the core saturation;
the pointing was dithered between exposures. All data were taken using an SDSS i$^{\prime}$ filter.

\subsection{MiniMo}

Observations with MiniMo were made on the nights of September 26-27, 2011 and 
February 17-18, 2012. The mosaic image consists of two $2048\times4096$ chips, 
with a plate scale of 0.141$^{\prime\prime}$/pixel, separated by a 7.8$^{\prime\prime}$ gap.
This gives a field of 9.6 arcmin$\times$9.6 arcmin.

\subsection{SOI}

Observations with SOI were made on the nights of March 21-23, 2012. The SOI mosaic also has 
two $2048\times4096$ chips, again split by a 7.12$^{\prime\prime}$ gap. Seeing was 
only moderate quality during the run, so we observed with 2x2 binning, for a
plate scale of 0.153$^{\prime\prime}$/binned pixel and a 5.2 arcmin$\times$5.2 arcmin FoV. We targeted
5 dithered exposures for each object. We generally suffered significant core saturation
for the brighter BL Lacs.

\begin{table*}[t!!]
\centering
\caption{Observations and Fitted Component Amplitudes and Errors}
\begin{tabular}{llrrrrrrrr}
Name&SIMBAD Name& Tel&Exp (s)&DIQ($^{\prime\prime}$)&$N_{PSF}$ &Core$^a$&Host$^a$&$\sigma_{stat}^a$&$\sigma_{sys}^a$\\
\hline
J0114+1325&  GB6 J0114+1325  &W&$3\times300$&0.72&26 &2160&144& 5.47&43.8\\
J0115+2519&  RX J0115.7+2519 &W&$3\times300$&0.79&21 &1050&377& 5.27&18.8\\
J0222+4302&  3C 66A          &W&$5\times300$&0.68&36&13500&229& 7.77&315.\\
J0316+0904&  GB6 J0316+0904  &W&$5\times300$&0.63&27 &1910&133& 4.18&61.5\\
J0543$-$5532&1RSX 053810.0$-$390839&S&$5\times300$&0.57&20 &3410&638& 6.82&108.\\
J0558$-$7459&PKS 0600-749    &S&$5\times180$&0.69&24 &1260&-24& 7.29&56.7\\
J0700$-$6610&PKS 0700-661    &S&$4\times300$&1.06&19 &4950& 81& 6.70&147.\\
J0721+7120&  S5 0716+71      &W&$600+3\times300$&1.34&19 &39700&5430&16.0&812.\\
J1023$-$4336&RX J1023.9$-$4336&S&$5\times300$&0.62&21 &8610&120& 10.3&106.\\
J1026$-$8543&PKS 1029$-$85   &S&$1\times300$&0.83&16 &2150& 10& 6.21&88.6\\
J1110$-$1835&CRATES J1110$-$1835&S&$5\times180$&0.61&9  & 372&  9& 3.84&20.0\\
\hline
\end{tabular}

\leftline{
Tabulated quantities: Name, Simbad Name, Telescope, exposure used, final combined image
full width at half maximum, number of
}\leftline{
PSF stars used, fit core (PSF) count rate, fit host
count rate, statistical error on host rate, estimated systematic error on host rate.
}

\leftline{$^a$ counts/s.}
\end{table*}
\bigskip

\subsection{Data Processing}

All images were reduced using the IRAF mscred package for mosaic image data. Standard zero image
bias subtraction and dome flats were applied and cosmic rays were cleaned using the 
IRAF xzap package. A few bad pixel/cosmic ray events were edited by hand. The $i^\prime$ fringing
was quite modest, especially for the chip hosting the BL Lac target,
so no fringe corrections were made. 

	After generation of a world coordinate system (WCS) for each frame using the 
USNO B-1 reference catalog, the dithered frames were stacked to a median combined frame. 
In general, the PSF varied relatively slowly during the observations.
In a few cases, we rejected the worst sub-frames before the final image combination.
The final exposures included in the image stack and the stacked image
FWHM are listed in Table 1.

\section{Image Modeling}

	Our goal is to measure the unresolved AGN core and surrounding resolved 
host galaxy, extracting reliable fluxes or flux upper limits for the latter. To 
this end we model cut-outs around each BL Lac. Typically we treated
a $10^{\prime\prime}\times10^{\prime\prime}$ region, although for 1 object
we used a $22^{\prime\prime}\times22^{\prime\prime}$ region to contain the bulk of the
host counts, and for 2 other objects we were able to reduce the region size 
to $6^{\prime\prime}\times6^{\prime\prime}$ while containing all the host flux.

\subsection{Model Components}

	Since we are interested in the faint host excess in the wings of the AGN PSF, we need
an accurate PSF for each final combined image frame. These PSFs were generated using the 
IRAF daophot package. Generally over 20 bright isolated stars were available to generate the
PSF, although a few fields were more sparse. Saturated stars were not included in the PSF stack.
For most cases our PSF model extends to 5$^{\prime\prime}$, where the host contribution drops 
well below the sky, however for 2 objects with the best PSF we used 3$^{\prime\prime}$ and 
for the target with the worst seeing we extended the PSF model to 11$^{\prime\prime}$. Since 
we needed much of the field to include sufficient bright stars, we used a quadratic 
variation across the image for the analytic PSF core. The accuracy of the PSF model
was checked by generating a model (using the quadratic position-dependent PSF) for the precise
positions of a set of check stars in each image. The residuals after subtraction were very small
well out into the PSF wings, although as expected, poor subtraction was often present for
near-saturated cores within 1 FWHM. The target PSF model was generated for the field 
position of the BL Lac.

Based on previous studies of BL Lac host galaxies \citep{scarpa00,urry00}, we assume 
that our hosts are well modeled by a de Vaucouleurs profile of Sersic index 4. In
addition to the integrated model flux we have up to three shape parameters. For bright,
well resolved hosts, we can fit the effective angular size $\theta_e$. If not fit, we fix this at
$\theta_e=1.65^{\prime\prime}$, which corresponds to R$_e=$10 kpc at a typical BL Lac $z=0.5$ for 
our standard approximate concordance cosmology ($\Omega_m=0.3$, $\Omega_{\Lambda}=0.7$,
$H_0=70$km\,s$^{-1}$ Mpc$^{-1}$). Occasionally resolved hosts show significant ellipticity;
we then fit this along with the position angle. Before fitting the model profile
is convolved with the locally generated model PSF for the particular image.

\subsection{Fitting Procedure}

	In all cases, we use the AGN core position (fit with IRAF DAOPHOT) determined at sub-pixel
accuracy to generate the normalized templates for the PSF and PSF-convolved host models.
The core and host position were not adjusted in the fit. Before fitting we masked saturated
pixels (counts $> 35,000$DN) which typically excluded $\sim 1$FWHM around the PSF core
and for the brightest sources, a modest number of pixels in a bleed trail. We also had the
option of masking pixels associated with neighboring sources (companion galaxies and field
stars). This was done for 3 sources.

We fit the masked cut-out images with $\chi^2$ minimization of the residuals to the model
counts in each image pixel, using a Nelder-Mead downhill simplex algorithm \citep{scipy1}. 
Host shape parameters are determined hierarchically. If a statistically
significant amplitude is fit ($> 3 \sigma_{\rm stat}$) using the default spherical
host with fixed $\theta_e$, we re-fit allowing $\theta_e$ to vary. If $\theta_e$ 
is measured with high statistical 
significance, we re-fit including the ellipticity and position angle of the host.
In this data set only three sources had a well measured
effective radius and only one had significant ellipticity.
The routine returns best fit values for the PSF, host and constant background counts,
up to three additional host shape parameters, and statistical errors on each fit
quantity.

\subsection{Systematic Errors}

\begin{figure}[t!!]
\vskip 8.3truecm
\includegraphics{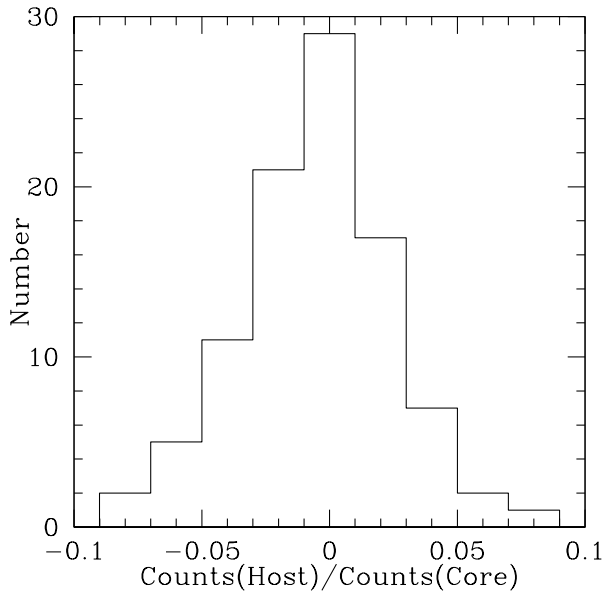}
\caption{\label{systdist}
Distribution of fit 'host' counts to PSF counts for fits to isolated stars.
The symmetric distribution centered on zero shows that, across all fields, there
is little overall systematic bias in the PSF model.}
\end{figure}

Inevitably, due to peculiarities of the actual image and imperfections in the model PSF,
we expect systematic errors to dominate the simple fit statistical errors. To estimate 
the systematic errors on the host fitting, we selected
$\sim$10 bright stars in each image, generated the local PSF and fit for combined PSF,
de Vaucouleurs (fixed $\theta_e$) `host' and uniform background counts. Ideally these would be
not be drawn from the PSF stars, but we did not always have enough bright stars to avoid
overlap. The fit `host' counts give an estimate of the systematic host errors.
Figure \ref{systdist} shows the distribution of fit `host' counts, as a fraction of the 
stellar (PSF) counts. The mean of zero suggests good model PSFs with no overall bias.

\begin{figure}[t!!]
\vskip 8.5truecm
\includegraphics{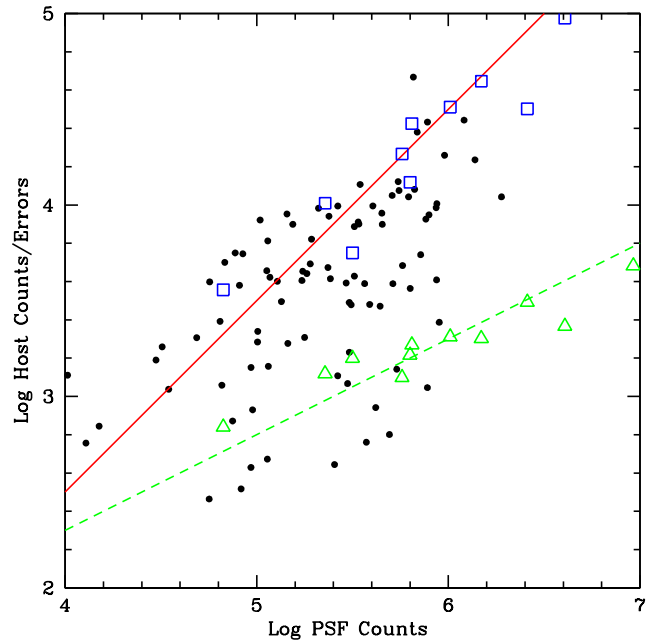}
\caption{\label{fiterr}
Here black points show the distribution of the fit `host' counts fit for isolated stars,
plotted as a function of the PSF counts. Open triangles show the statistical
errors to the BL Lac host fits, while open squares show our estimate of the
systematic host error for each BL Lac field (see text).}
\end{figure}

	In figure \ref{fiterr} we show the absolute value of the individual fit 
`host' counts plotted as a function of PSF counts. If the errors were purely statistical
we would expect a square-root scaling, instead the upper envelope of the distribution
is approximately linear (solid line), suggesting that for bright PSFs the dominant errors are indeed
systematic. The figure also show the statistical fit errors for the individual BL Lacs,
with the expected $\sim$square root trend (dashed line). Note that the distribution of systematic
`host' amplitudes fit to stars lies above this trend, with statistical errors significant
when the PSF contains less than $\sim 10^4$ counts and systematics increasingly dominating
for brighter stars.

	Since the systematic errors grow linearly with PSF flux, for each BL Lac
field we normalize the stellar `host' fits to the PSF flux, take the RMS of this
distribution as $f_{rms}$ and compute $\sigma_{sys}= f_{rms} \times N_{AGN}$, where 
$N_{AGN}$ is the fit core counts, to be our estimate of the systematic error in
our host count estimate. These errors (Figure \ref{fiterr}) lie $\sim 5- 10\times$ 
above the $\sigma_{\rm stat}$ fitting errors for the corresponding field.
We believe that this estimate is conservative, as the resulting errors lie 
at the upper envelope of the individual stellar `host' errors.

\begin{figure*}[ht!!]
\vskip 8.5truecm
\includegraphics{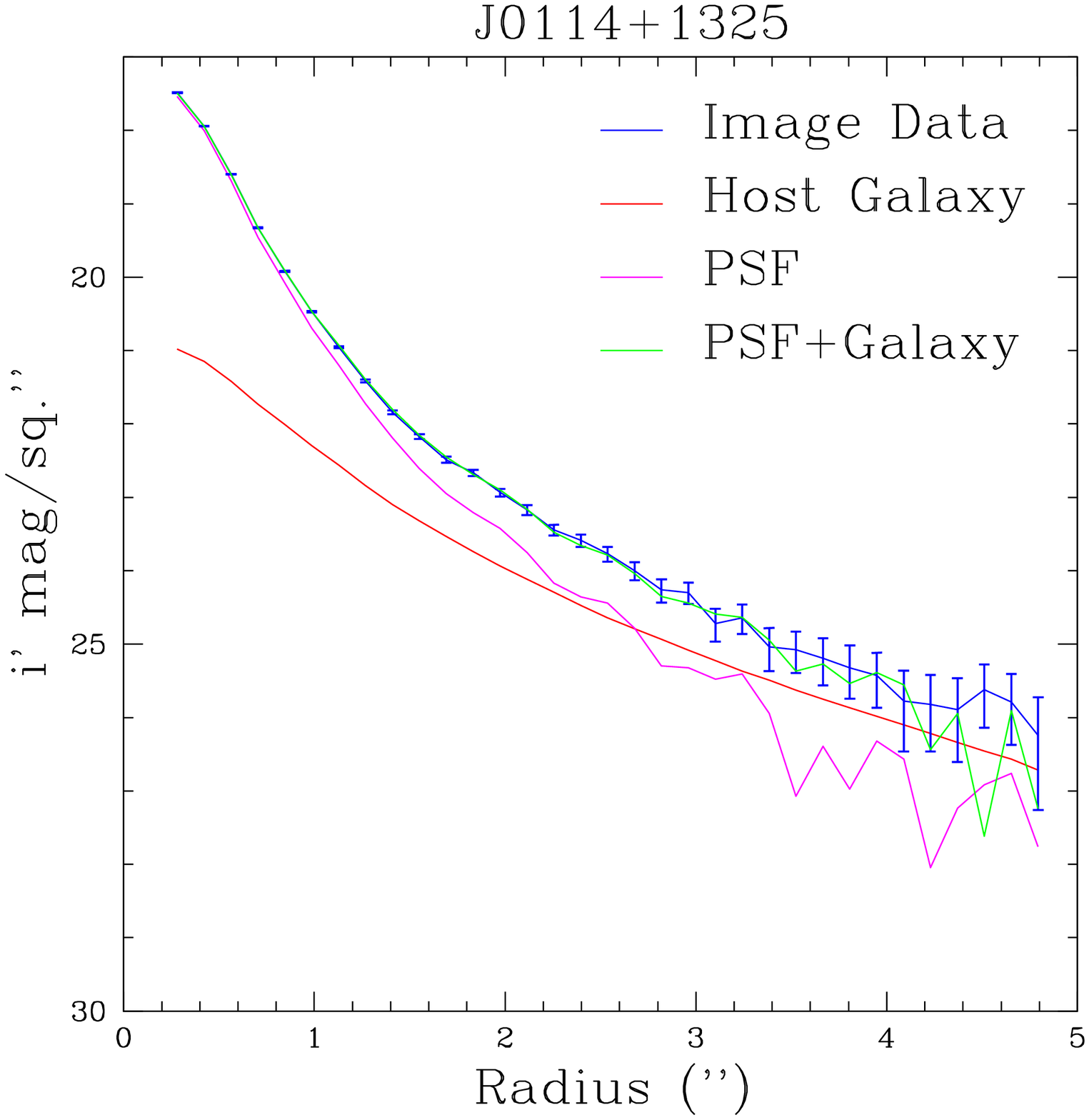}
\includegraphics{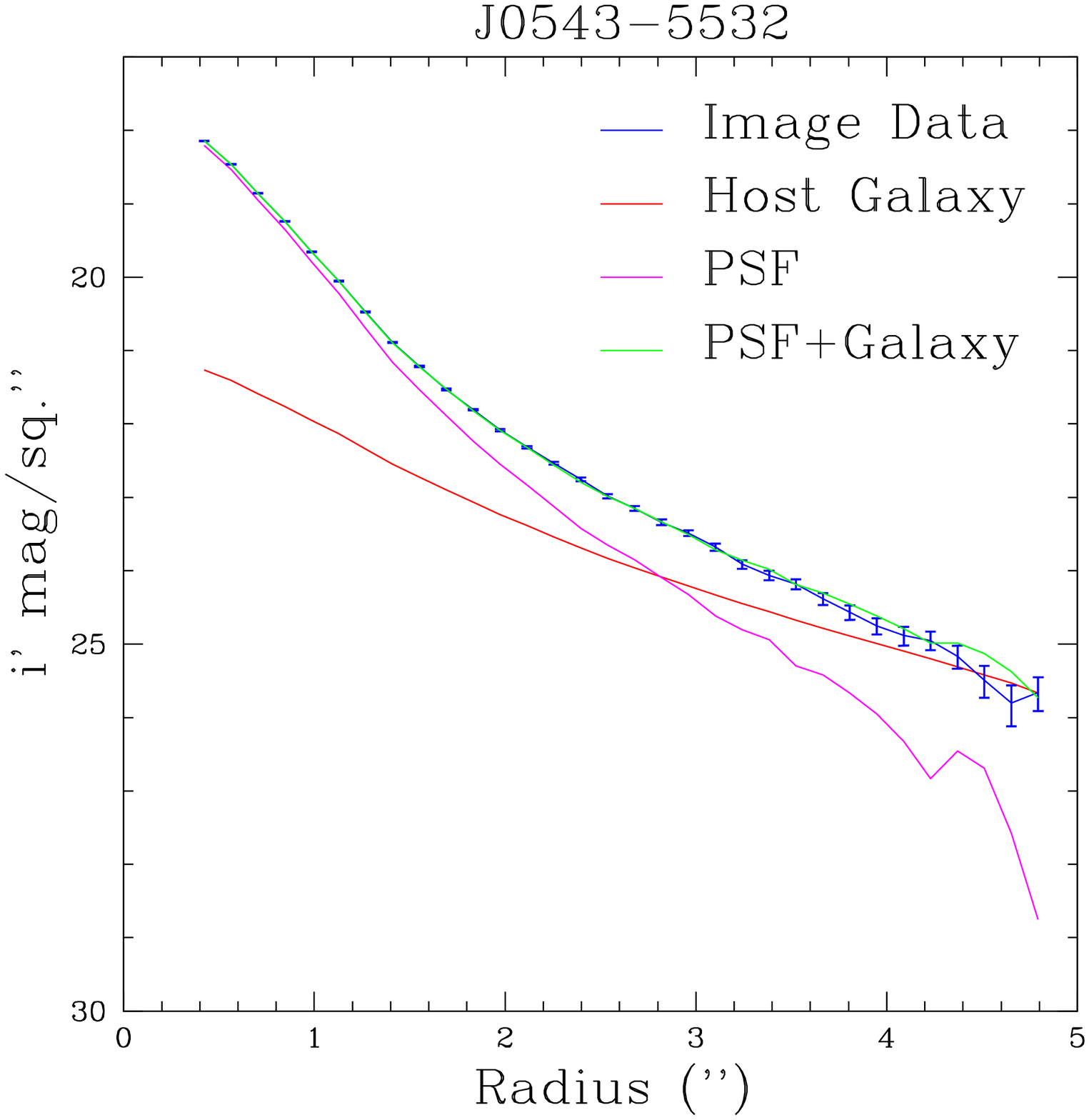}
\caption{\label{Detections} 
Two objects with significant host detections. Left: J0114+1325 (WIYN), Right: J0543-5532 (SOAR).
The inferred redshifts are $z\approx 0.58$ and $z\approx 0.37$, respectively. The lines
show the profile of the best-fit model components, converted to magnitudes/arcsec$^2$.
For both the host counts dominate the PSF beyond $\sim 2.7^{\prime\prime}$.}
\end{figure*}

\subsection{Fitting Results}

	The final fit component counts and the statistical and systematic
errors are listed in Table 1.
We also list the calibrated core magnitudes and host flux ratios in Table 2,
where we convert $i^\prime$ image counts to
instrumental band flux using calibration stars from the SDSS.

	To claim a significant host detection, we require that the fit counts
exceed  $\sqrt{3\sigma_{stat}^2 + \sigma_{sys}^2}$. Four of our BL Lac hosts
are detected at very high significance. Two are seen with somewhat lower confidence,
with J0316+0904 having $\sim 2.2 \sigma_{sys}$ significance  and J1023$-$4336
appearing at $\sim 1.1 \sigma_{sys}$ (although both are of high statistical significance). 
Visual inspection of the images and the
azimuthally average radial profile plots (below) confirm that the former is a very
likely detection, but the latter, while a plausible detection, may be more safely treated 
as an upper limit.  When we have no significant host
detection, we infer host flux counts to be $< \sqrt{3\sigma_{stat}^2 + \sigma_{sys}^2}$.

	Only three sources have significant $R_e$ measurements listed. These
well measured sizes are all consistent with the standard 10\,kpc assumption. 
J0115+2519 was the only source with a statistically significant ellipticity $e=0.160\pm.043$;
this is at major axis position angle $-16.1\pm 3.5^\circ$, measured North through East.

\begin{figure*}[ht!!]
\vskip 8.5truecm
\includegraphics{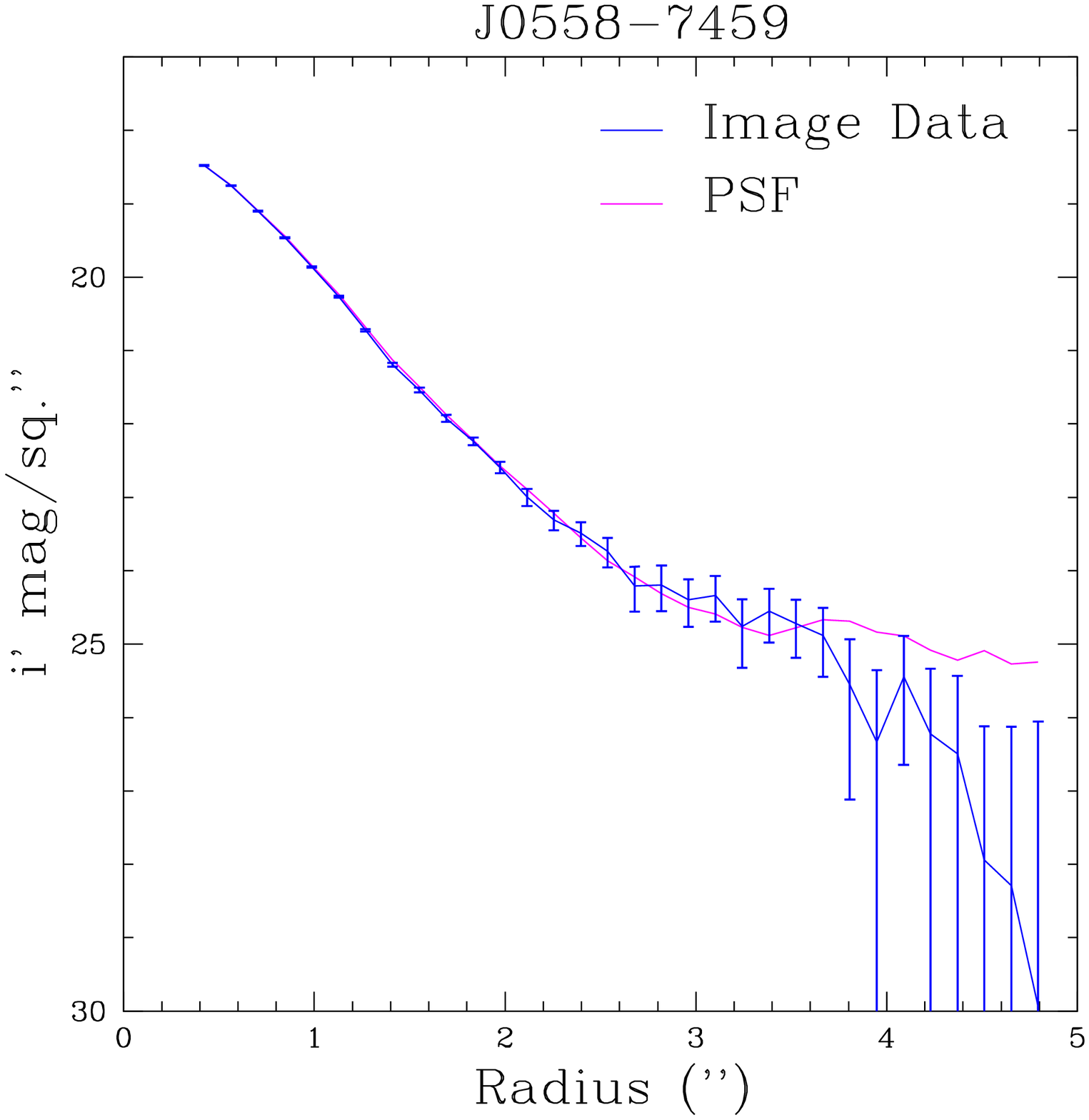}
\includegraphics{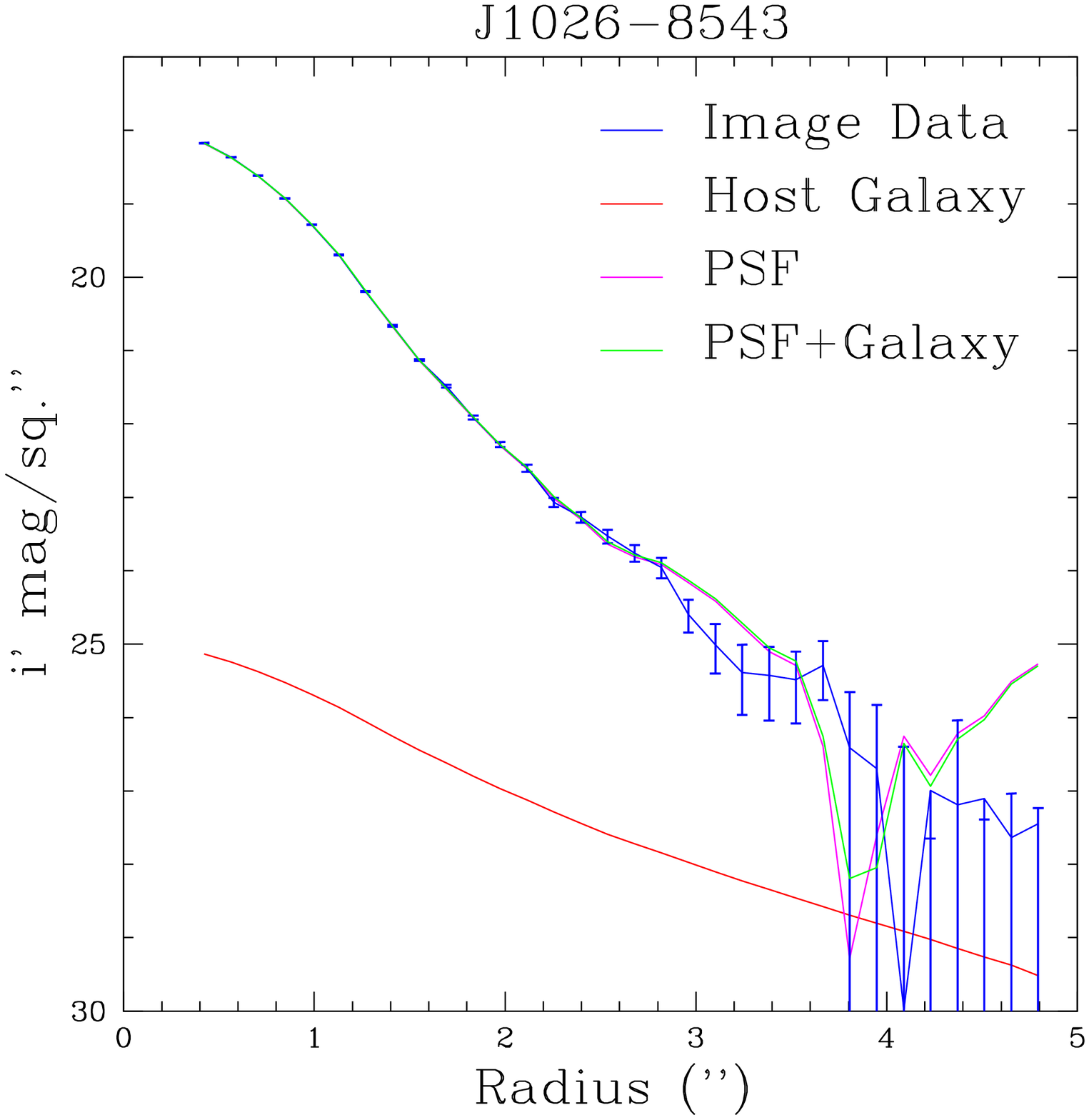}
\caption{\label{Limits} 
Two objects providing host upper limits. Left: J0558$-$7459 whose fit amplitude is
in fact negative, so no host model is shown; the inferred limit is $z>0.66$.
Right: J1026$-$8543 has an insignificant host detection. We infer $z>0.62$}
\end{figure*}

	It is convenient to display these fits as azimuthally averaged profiles
of the surface brightness. In figure \ref{Detections} we show two sources with 
well-detected hosts. Figure \ref{Limits} shows two sources with radial profiles
well matched to the PSF, resulting in host upper limits. Error flags on the data
curves show the $1\sigma$ fluctuations assuming Poisson statistics.

\section{Redshift Estimates and Conclusions}

	A primary goal of this exercise is to use the host fluxes to constrain the source
redshift, adopting the standard candle hypothesis. To this end we use the Hubble diagram curve
computed in \citet{meisner1}. This curve follows an elliptical host formed at $z=2$ 
and evolving according to the \citet{fr97} models to the observed $z$ in our standard cosmology,
computing the observed flux by folding through the $i^\prime$ filter.
However, a recent spectral survey of {\it Fermi} BL Lacs \citep{shaw13a}
finds that the host luminosity of these $\gamma$-ray sources is 0.4 magnitudes fainter
than the $M_R$ = -22.9 $\pm$ 0.5 reported by \citet{sba05}. We thus amend the Hubble diagram
by normalizing the evolving models to this decreased luminosity at $z=0$.

\begin{figure} [h!!]
\begin{center}
\includegraphics[width=9.5cm]{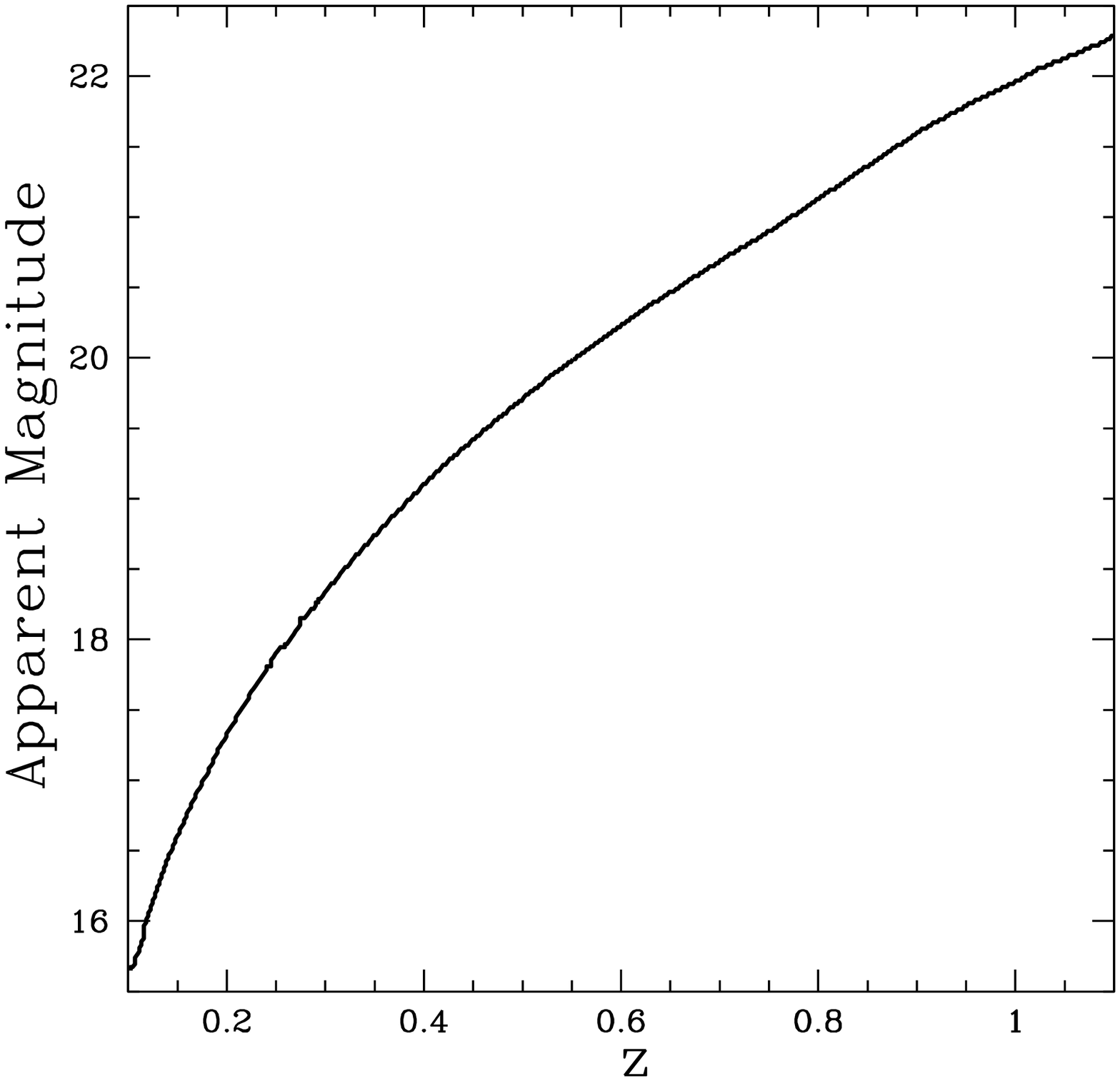}
\end{center}
\caption{$i^\prime$ Hubble diagram for gamma-ray BL Lac hosts, after \citet{meisner1}, but
normalized to a standard candle magnitude $M_R =-22.5$.}
\end{figure}

	Using this curve we translate the host magnitude values and upper and lower statistical and
systematic errors to redshift estimates and error ranges. These are reported in Table 2.
When only an upper limit on the flux is available, this translated to a redshift lower limit.
The redshift estimates for detected hosts varied from 0.13 to 0.58, with lower limits $0.42-0.93$.
We consider these limits conservative, in the sense that we use the revised (less luminous)
standard candle calibration above. However, insofar as some BL Lacs undoubtedly have
substantially sub-luminous hosts, individual sources may indeed appear at lower $z$.

\begin{table*}[ht!!]
\caption{Fitting Results: Calibrated Magnitudes, Sizes and Redshifts}
\centering
\begin{tabular}{lrrlrrll}
Name &SED$^a$&i$^{\prime}$$_{nucl}$&i$^{\prime}$$_{Host}$&$f_{host}/f_{nucl}$ &R$_e$ (kpc)&$z_{im}$&$z_{sp}$\\
\hline
J0114+1325& H &17.21&20.15$_{-0.037-0.29}^{+0.038+0.41}$  &0.07&8.32$\pm$0.14&0.583$_{-0.006-0.052}^{+0.007+0.064}$ &$0.61-1.63$\\
J0115+2519& H &17.68&18.80$_{-0.015-0.05}^{+0.015+0.05}$  &0.36&12.53$\pm$0.55&0.358$_{-0.001-0.006}^{+0.001+0.007}$&$0.37-1.63(>0.268)$\\
J0222+4302& I &15.13&$>$19.21                        &$<$0.02&-&$>$0.42                                            &$0.12-1.67$\\
J0316+0904& H &16.01&18.91$_{-0.033-0.412}^{+0.035+0.672}$&0.07&-&0.372$_{-0.004-0.076}^{+0.005+0.103}$              &$0.12-1.66$\\
J0543$-$5532& H &17.10&18.92$_{-0.012-0.17}^{+0.012+0.20}$&0.19&-&0.374$_{-0.002-0.023}^{+0.003+0.027}$             &$0.27-2.57$[=0.273]\\
J0558$-$7459& - &17.23&$>$20.52                        &$<$0.05&-&$>$0.66                                            &$0.29-2.20(>0.475)$\\
J0700$-$6610& I &16.28&$>$20.08                        &$<$0.03&-&$>$0.57                                            &$0.39-1.92$\\
J0721+7120& I &14.01&16.17$_{-0.0024-0.193}^{+0.0038+0.237}$&0.14&9.54$\pm$0.32&0.127$_{-0.001-0.073}^{+0.002+0.092}$             &$0.14-2.61$\\
J1023$-$4336& H &15.26&19.90$_{-0.090-0.689}^{+0.090+2.36}$&0.03&-&$^b$0.534$_{-0.014-0.115}^{+0.019+0.482}$             &$0.34-2.24$\\
J1026$-$8543& L &16.88&$>$20.31                        &$<$0.04&-&$>$0.62                                            &$0.32-2.30$\\
J1110$-$1835& L &18.69&$>$21.70                        &$<$0.06&-&$>$0.93                                            &$0.51-2.23$\\
\hline
\end{tabular}
\leftline{
Tabulated quantities: Name, SED class, Core magnitude, host magnitude and errors or limit, host radius,
}
\leftline{
inferred redshift or limit, spectroscopic constraints on redshift
}
\leftline{\footnotesize $^a$ SED class from \citet{2lac}, based on the synchrotron peak frequency.}
\leftline{\footnotesize $^b$ Good statistical, but marginal $\sim 1\sigma$ systematic significance. Corresponding lower limit is $z>0.42$}
\end{table*}

	\citet{shaw13a} analyzed high S/N spectra of a large number of {\it Fermi}
BL Lacs. In a method parallel to the present imaging host search, they detected
or placed limits on the BL Lac host by measuring the flux of a spectral component
from a standard candle elliptical. After appropriate k-correction and slit-loss corrections,
these measurements provided host redshift estimates or lower bounds. In addition,
sometimes intervening metal line absorption systems were detected. These provide firm,
model-independent lower bounds on the host redshift. These spectroscopically allowed ranges and lower
bounds are listed in the last column of Table 2; we give here the ranges derived
for a standard candle magnitude $M_R$ = -22.5, for consistency with our imaging results.

	Since in this program we targeted BL Lacs without known redshift,
none of our sources has a spectroscopic $z$ in \citet{shaw13a}.
For the five sources with host detection, two (J0316+0904 and J0543$-$5532)
have imaging redshift estimates consistent with the spectra-derived
bounds in that paper; the low significance detection of J1023$-$4336 is also consistent.
Three (J0114+1325, J0115+2519, and J0721+7120) lie at slightly lower $z$ than 
the spectral lower bound, but are within 3$\sigma$, and J0115+2519 is fully consistent 
with the strict lower limit provided by an intervening absorber. However, since completing
this study \citet{pita13} have published new high quality VLT/X-shooter spectra of
a number of blazars, including J0543$-$5532, measured here. For this source, they infer
$z=0.237$ based on a weak Ca II H/K doublet and a Na I absorption line. This redshift,
near the lower bound of  \citet{shaw13a} is well below our imaging estimate,
implying a host that is substantially fainter than our standard candle assumption.
At this $z$ our measured host flux implies an absolute magnitude 
$M_R = -21.67^{+0.01+0.20}_{-0.01-0.17}$, $0.8$mag ($1.7\sigma$) away from our assumed
standard candle luminosity. The host absolute magnitude is consistent with, but more 
accurate than, the spectroscopically estimated value in \citet{pita13}. 

	In six cases we derive lower bounds on the redshift; J1023$-$4336 may be interpreted
as a lower bound of $z> 0.42$.  These 
limits are always more constraining than those extracted from the spectroscopic
study. For J0558$-$7459 our new bound is also stronger than that obtained from the
intervening absorption line system. Thus these bounds may be useful in BL Lac
population studies \citep[e.g.][]{aet14}.

	With median $f_{host}/f_{nucl} = 0.07$ our sources are very strongly dominated by
the non-thermal nuclear core flux. This is in contrast to the HST study of \citet{scarpa00},
where of 69 BL Lacs with resolved hosts, 37 had $f_{host}/f_{nucl}>1$. Our large
core dominance is similar to that found in \citet{meisner1} and, as noted there,
it may be attributed to the fact that these are gamma-ray selected BL Lacs and thus
should have high alignment  between the jet axis and the Earth line-of-sight,
increasing the core dominance. In addition, these sources were drawn from the BL Lacs
lacking redshifts even after extensive spectroscopy with 8\,m-class facilities
\citep{shaw13a,shaw13b}. Since sources with brighter hosts allow easier absorption line
redshift measurements, the remainder (including the sources studied here) should
have an especially high core dominance.

	In previous studies \citep{scarpa00,urry00,meisner1} the BL Lacs were seen to 
have an excess of faint galactic companions, indicating that they were located in cluster 
environments and that they may have recent interaction activity. In general, the relatively
poor seeing and modest image depth achieved during this project prevented us from identifying
very faint companions. For 2 objects, we did find bright companion galaxies nearby. 
J0114+1325 had two companions within a 3$^{\prime\prime}$ ($\sim 20$\,kpc) radius. 
These companions were close enough that they overlap significantly with the BL 
Lac host wings, making accurate flux measurement difficult; we found their i$^{\prime}$ 
magnitudes to be roughly 20.5 and 20.8. J0222+4302 had 4 surrounding bright companions 
located 11$-12^{\prime\prime}$ ($\sim 70$\,kpc) from the core, 
with i$^{\prime}$ magnitudes of 18.9, 19.3, 19.5, and 20.9.

\bigskip
\bigskip

	We have detected hosts in the images of 5/11 BL Lacs, plus one marginal detection. 
Assuming a
standard candle host luminosity these provided redshift estimates $z=0.127 - 0.583$
($z_{med}$ = 0.37). The minimum redshifts for the remainder are also larger
than available spectroscopic lower limits and so are useful, as well.
Perhaps unsurprisingly, our host detections were for four HBLs (high spectral
peak energy, relatively low luminosity) sources and one intermediate peak source.
The other intermediate peak sources and the two LBLs (low peak, higher power) sources
in our sample yielded only lower limits on $z$; these include the highest limit,
$z>0.93$ in this sample.

When no other redshift estimate is available, these imaging-derived values
can be useful for statistical purposes, e.g. in population studies. However, as
emphasized by the spectroscopic $z$ recently derived for J0543$-$5532, the standard 
candle hypothesis is only statistically useful, at best, and should be subject to
further study. Indeed \citet{pita13} estimate that two of their BL Lac hosts
have $M_R < -24$, even further from the expected standard candle value. Our host
flux measurements thus remain useful whenever an independent redshift is derived.
The prospects for further spectroscopic redshifts of our imaged hosts are, in fact,
good; these are excellent targets for spatially resolved spectroscopy, 
especially with Integral Field Unit (IFU) feeds, which can isolate host spectral
features from the wings of the BL Lac.

\acknowledgements
	This work was supported in part by NASA grants NNX11AO44G
and NNX12AP85G and the Stanford Vice-Provost for Undergraduate research.
We thank Sasha Brownsberger for assistance with the data analysis and the
referee for a careful detailed reading.

\end{document}